\documentclass[twocolumn,letter]{jpsj3}

\usepackage{graphicx}
\usepackage{dcolumn}
\usepackage{bm}
\usepackage{color}
\usepackage{ulem}

\begin{document}

\title{Clear Reduction in Spin Susceptibility and Superconducting Spin Rotation for $H \parallel a$ in the Early-Stage Sample of Spin-Triplet Superconductor UTe$_2$}

\author{Shunsaku~Kitagawa$^{1,}$\thanks{E-mail address: kitagawa.shunsaku.8u@kyoto-u.ac.jp}, 
Kousuke~Nakanishi$^{1}$, 
Hiroki~Matsumura$^{1}$,
Yuki~Takahashi$^{1}$,
Kenji~Ishida$^{1}$, 
Yo~Tokunaga$^{2}$, 
Hironori~Sakai$^{2}$, 
Shinsaku~Kambe$^{2}$,
Ai~Nakamura$^{3}$,
Yusei~Shimizu$^{3}$,
Yoshiya~Homma$^{3}$,
Dexin~Li$^{3}$,
Fuminori~Honda$^{3,4}$,
Atsushi~Miyake$^{3}$, and
Dai~Aoki$^{3,5}$
}

\inst{$^1$Department of Physics, Graduate School of Science, Kyoto University, Kyoto 606-8502, Japan \\
$^2$Advanced Science Research Center, Japan Atomic Energy Agency, Tokai, Ibaraki 319-1195, Japan \\
$^3$Institute for Materials Research, Tohoku University, Oarai, Ibaraki 311-1313, Japan \\
$^4$Central Institute of Radioisotope Science and Safety, Kyushu University, Fukuoka 819-0395, Japan \\
$^5$Universit\'e Grenoble Alpes, CEA, IRIG, PHELIQS, F-38000 Grenoble, France
}

\date{\today}

\abst{
We report the re-measurement of the $a$-axis spin susceptibility component in an early-stage sample of the spin-triplet superconductor UTe$_2$ with the transition temperature of $T_{\rm SC}$ = 1.6 K.
Using Knight-shift measurements along the $b$ axis and at a 10-degree tilt from the $b$ axis towards the $a$ axis, we accurately determined the $a$-axis component without directly measuring the $a$-axis Knight shift.
Our results reveal a decrease of approximately 3\% in the $a$-axis spin susceptibility in the superconducting state under $a$-axis magnetic field $\mu_0 H_a \sim 0.1$ T, indicating that the spin susceptibility decreases similarly in both early-stage and ultraclean samples with $T_{\rm SC}$ = 2.1 K.
The previously reported absence of the reduction in Knight shift is attributed to the missing of signal from the superconducting region and to the detection of residual signals from the non-superconducting region instead.
We also found that the decrease in the $a$-axis spin susceptibility is immediately suppressed with increasing the $a$-axis magnetic field and is estimated to be completely suppressed at around 1.5 T due to superconducting spin rotation.
}

\maketitle

Superconductivity is a quantum condensed state arising from the formation of Cooper pairs by two electrons\cite{J.Bardeen_PR_1957}.
In many cases, including heavy-fermion superconductors\cite{H.Tou_JPSJ_2005,S.Kitagawa_PRL_2023} and high-temperature cuprate superconductors\cite{J.Nachtigal_condmatt_2020}, the pair typically forms with antiparallel spins, resulting in a spin-singlet state without spin degrees of freedom\cite{D.Khim_JPSJ_2023}.
In contrast, a spin-triplet state with spin degrees of freedom, where the electron spins are parallel, is also possible and has been realized in superfluid $^3$He\cite{A.J.Leggett_RevModPhys_1975}. 
The spin-triplet superconducting state introduces qualitatively different superconducting properties due to its spin degrees of freedom, such as superconducting spin rotation\cite{K.Machida_JPSJ_1998} and peculiar spin excitation\cite{J.B.Ketterson_PRB_1992}, which are absent in the spin-singlet superconductors.
However, the number of candidate materials for spin-triplet superconductivity is less than ten\cite{H.Tou_PRL_1996,S.Saxena_Nature_2000,K.Ishida_PRL_2002_2,K.Matano_NatPhys_2016,J.Yang_SciAdv_2021,S.Ogawa_JPSJ_2023,D.Aoki_JPSJ_2019}.
In addition, these few candidate materials often coexist with magnetic phases or exhibit very small spin susceptibility in the normal state, leaving many aspects of their superconducting properties and impurity effects experimentally unclear.

\begin{figure}[tb]
\centering
\includegraphics[width=0.5\textwidth]{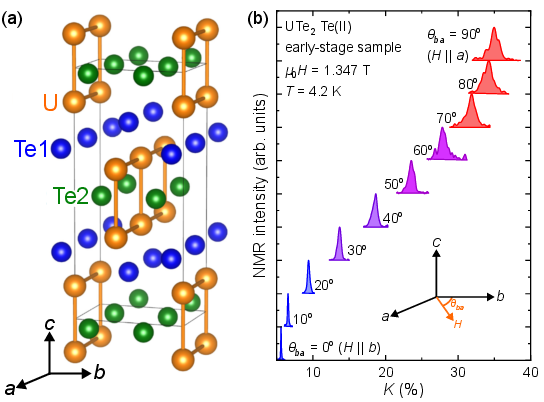}
\caption{(Color online) (a) Crystal structure of UTe$_2$ drawn by VESTA\cite{K.Momma_JAC_2011}. 
(b) $^{125}$Te-NMR spectrum against the angle between the crystalline axis and a magnetic field in the $ab$ plane measured at 4 K.
The $\theta_{ba}$ is defined as the angle from the $b$ axis to the $a$ axis.
}
\label{fig.1}
\end{figure}

\begin{figure*}[tb]
\centering
\includegraphics[width=\textwidth]{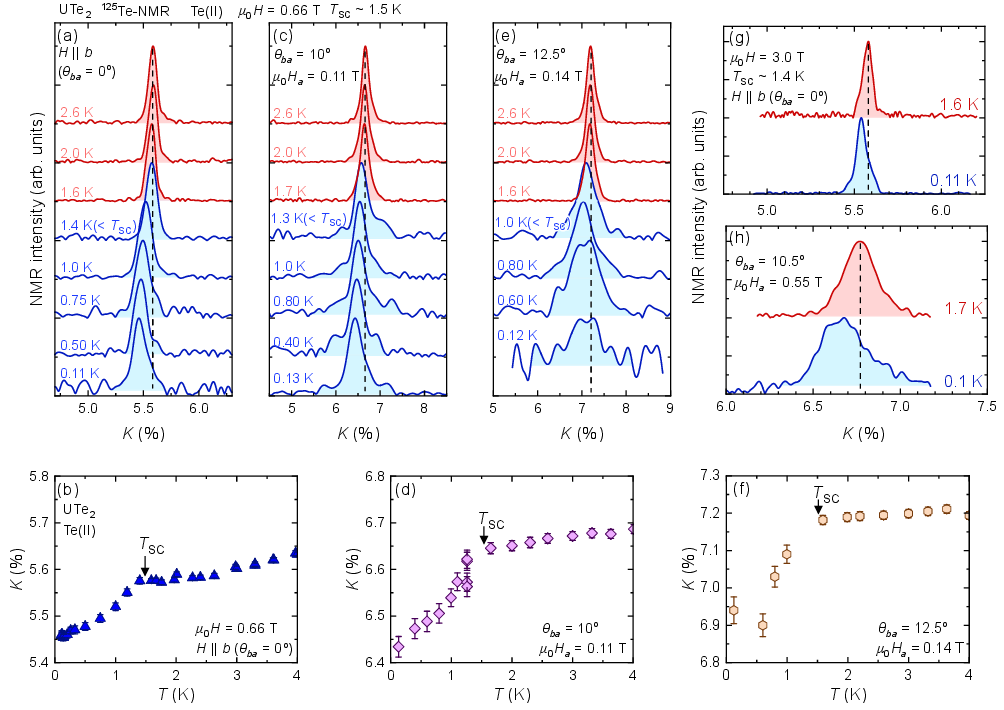}
\caption{
(Color online) Temperature variation of the $^{125}$Te-NMR spectra and the temperature dependence of the Knight shift for $H \parallel b$[(a),(b)], the magnetic field rotated by 10 degrees towards the $a$ axis ($\theta_{ba} = 10^{\circ}$)[(c),(d)], and $\theta_{ba} = 12.5^{\circ}$[(e),(f)] measured at 0.66~T.
The solid arrows indicate $T_{\rm SC}$.
$^{125}$Te-NMR spectrum above and below $T_{\rm SC}$ for $H \parallel b$ (g) and $\theta_{ba} = 10.5^{\circ}$(h).
The intensity of each spectrum is normalized by the peak value.
The broken lines indicate Knight shift in the normal state.
}
\label{fig.2}
\end{figure*}

In this context, we focus on the recently discovered uranium-based superconductor UTe$_2$\cite{S.Ran_Science_2019,D.Aoki_JPSJ_2019_2,D.Aoki_JPCondMatt_2022}.
UTe$_2$ crystallizes in an orthorhombic structure with space group $Immm$ ($D_{2h}^{25}$, \#71) and features a two-leg ladder structure along the $a$ axis composed of uranium atoms, as shown in Fig.~\ref{fig.1}(a).
Initially discovered with a superconducting transition temperature $T_{\rm SC}$ of 1.6~K, improvements in synthesis methods have enabled the ultraclean single crystals without uranium deficiencies, raising $T_{\rm SC}$ to 2.1~K\cite{H.Sakai_PRM_2022,D.Aoki_JPSJ_2024}.
Since its discovery, UTe$_2$ has been a promising candidate for spin-triplet superconductors due to its large upper critical field $H_{\rm c2}$\cite{S.Ran_Science_2019,G.Knebel_JPSJ_2019,G.Knebel_JPSJ_2020} and field/pressure-induced superconducting phases\cite{A.Rosuel_PRX_2023,D.Braithwaite_CommunPhys_2019,D.Aoki_JPSJ_2020,S.Ran_NatPhys_2019,W.Knafo_CommPhys_2021,H.Sakai_PRL_2023,F.Honda_JPSJ_2023,M.Ajeesh_JPSJ_2024}.
Our nuclear magnetic resonance (NMR) measurements of the spin susceptibility provide further evidence suggesting a spin-triplet state in UTe$_2$\cite{G.Nakamine_JPSJ_2019,G.Nakamine_PRB_2021,G.Nakamine_JPSJ_2021,H.Fujibayashi_JPSJ_2022,H.Matsumura_JPSJ_2023,K.Kinjo_PRB_2023,K.Kinjo_SciAdv_2023}. 
However, discrepancies between results from the early-stage and ultraclean samples are present.
Although both early-stage and ultraclean samples exhibit a slight decrease in the spin susceptibility in the superconducting state for $H \parallel b$ and $H \parallel c$, the ultraclean sample shows a significant decrease in the spin susceptibility for $H \parallel a$\cite{H.Matsumura_JPSJ_2023}, whereas such a decrease was not detected in the early-stage sample\cite{H.Fujibayashi_JPSJ_2022}.
As the decrease in the spin susceptibility indicates that the $d$ vector component, which is the order parameter in spin-triplet superconductors, is in the magnetic field direction, the difference in results for $H \parallel a$ suggests a difference in superconducting symmetry between the samples.
The difference may arise from (i) the possible rotation of the superconducting spin under small magnetic fields, (ii) suppression of the spin susceptibility reduction due to minor impurities and defects in the samples, or (iii) the observation of the signals from the non-superconducting region.

In this work, we performed a re-evaluation using NMR measurements in the early-stage UTe$_2$ sample to clarify whether the reduction in the $a$-axis spin susceptibility in the superconducting state is intrinsic or not.
Specifically, we measured the spin susceptibility under magnetic fields applied along the $b$ axis and at an angle of approximately 10 degrees from the $b$ axis towards the $a$ axis.
As the $a$ axis is the magnetic easy axis and has a much larger value of Knight shift than the $b$ axis\cite{Y.Tokunaga_JPSJ_2019}, it is possible to measure the $a$-axis spin susceptibility component with a small projection field along the $a$ axis.
Accurate measurements of the spin susceptibility along the $a$ axis have been challenging because the $a$-axis magnetic field causes additional linewidth broadening in the early-stage sample\cite{Y.Tokunaga_JPSJ_2022,H.Fujibayashi_JPSJ_2023} and the signal intensity is quite weak\cite{H.Fujibayashi_JPSJ_2022}, as shown in Fig.~\ref{fig.1}(b).
However, by employing a technique utilizing the projection components, we achieved precise measurements of the spin susceptibility along the $a$ axis.
Our results reveal that even in the early-stage sample, the spin susceptibility along the $a$ axis decreases in the superconducting state.
Furthermore, we observed the suppression of this decrease under a very small magnetic field of approximately 1.5 T well below $\mu_0 H_{\rm c2} = 7$ T, indicating a rotation of the superconducting spin to the applied field direction.

The $^{125}$Te (nuclear spin $I = 1/2$, gyromagnetic ratio $^{125}\gamma/2\pi = 13.454$ MHz/T)-NMR measurements were performed on a single crystal of dimensions $1.5 \times 1.6 \times 2.6$ mm$^3$, identical to the sample used in our previous study\cite{H.Fujibayashi_JPSJ_2022}. 
The sample was grown using the chemical vapor transport method \cite{D.Aoki_JPSJ_2019_2}. 
To get strong signal intensity, $^{125}$Te is enriched to 99.9\%.
$T_{\rm SC}$ = 1.6 K, was determined by ac susceptibility measurements using an NMR coil.
UTe$_2$ exhibits two distinct $^{125}$Te-NMR signals, originating from the two inequivalent crystallographic Te sites, as shown in Fig.~\ref{fig.1}(a).
We measured the Knight shift in the $^{125}$Te-NMR signal with the larger Knight shift [Te(II)].
The frequency-swept NMR spectra were acquired by performing the Fourier transform of the spin-echo signal observed after a spin-echo radiofrequency (RF) pulse sequence in a fixed magnetic field.
The NMR spectra are presented with the horizontal axis representing $K = (f - f_0)/f_0$.
Here, $f_0 = (^{125}\gamma/2\pi)\mu_0H$.
The magnetic field was calibrated using a $^{65}$Cu ($^{65}\gamma/2\pi = 12.089$ MHz/T)-NMR signal with the Knight shift $K_{\rm Cu} = 0.2385$\% from the NMR coil\cite{MetallicShift}.
The Knight shift was determined by the peak position of the NMR spectrum.
The NMR spectra in the superconducting state were recorded following a field-cooling process to avoid random vortex configuration.
The sample was rotated within the $ab$ planes using a split-pair magnet equipped with a single-axis rotator.
By examining the magnetic field-angle dependence of the Knight shift, we precisely aligned the sample along the $b$ axis with an angular resolution of $0.5^\circ$.
For the low-temperature NMR measurements down to 0.1 K, the crystal with the NMR coil was immersed in a $^3$He/$^4$He mixture.
The energy of the RF pulses was reduced to ensure that the NMR results remained unaffected by the power of the RF pulses\cite{K.Ishida_JPSJ_2020}.

We first present the results for the spin susceptibility along the $b$ axis.
Figure~\ref{fig.2}(a) shows the temperature variation of the NMR spectra for $H \parallel b = 0.66$ T.
As the temperature decreases, the spectra gradually shift to a lower Knight shift ($K$) side.
A significant reduction in the intensity with linewidth broadening was observed below $T_{\text{SC}}$.
The reduction in the intensity and increase in linewidth can be attributed to the superconducting diamagnetism and/or distribution in susceptibility.
The temperature dependence of the Knight shift derived from the peak values is shown in Fig.~\ref{fig.2}(b).
The anomalous broadening below 1~K reported in the previous measurements\cite{G.Nakamine_JPSJ_2021} was not observed in this sample.
The decrease in Knight shift in the superconducting state is estimated to be 0.12\%.
The experimental results for $H \parallel b$ are consistent with our previous studies\cite{G.Nakamine_JPSJ_2019,G.Nakamine_PRB_2021}.

Next, the same measurements were conducted with the magnetic field tilted towards the $a$ axis.
Figure~\ref{fig.2}(c) presents the NMR spectra obtained with the magnetic field rotated by 10 degrees towards the $a$ axis ($\theta_{ba} = 10^\circ$).
The magnetic field strength is 0.66 T, the same as for $H \parallel b$, whereas the projected magnetic field along the $a$ axis is $\mu_0 H_a = \mu_0 H\sin10^\circ = 0.11$~T.
The $a$-axis component makes the linewidth broader even in the normal state; however, a clear shift to lower $K$ values was observed below $T_{\text{SC}}$.
The temperature dependence of the Knight shift, shown in Fig.~\ref{fig.2}(d), indicates a decrease of 0.21\% in the superconducting state, which is roughly twice larger than the decrease observed along the $b$ axis.
The NMR spectrum at a further tilted angle of $\theta_{ba} = 12.5^\circ$ is shown in Fig.~\ref{fig.2}(e).
A further rotation of 2.5$^\circ$ broadened the NMR spectrum significantly, and a clear double-peak structure was observed below 0.60 K.
This behavior is likely due to the presence of regions within the sample that remain non-superconducting, possibly induced by a tiny amount of U deficiency\cite{Y.Haga_JPCM_2022}.
Tilting the magnetic field towards the $a$-axis weakens the signal from the superconducting regions, resulting in the observed double-peak structure.
Even in the spectra for $H \parallel b$ and with $\theta_{ba} = 10^\circ$, there is residual spectral weight near the normal state position.
This is much weaker in the ultraclean sample\cite{H.Matsumura_JPSJ_2023}.
With further rotation, the signal in the superconducting state could no longer be observed in this study.
The temperature dependence of lower $K$ peak at $\theta_{ba} = 12.5^\circ$ is shown in Fig.~\ref{fig.2}(f).
As shown in Figs.~\ref{fig.2}(g) and \ref{fig.2}(h), similar measurements were performed at a magnetic field strength of 3~T, and the decreases in Knight shift in the superconducting state were found to be 0.04\% for $H \parallel b$ and 0.10\% for the 10.5$^\circ$ tilt.

Here, we determine the $a$-axis component of the spin susceptibility from the measured data.
Generally, when the field is rotated by an angle $\theta_{ba}$ from the $b$ axis towards the $a$ axis, the Knight shift at the angle $\theta_{ba}$ $K_\theta$ can be expressed as,
\begin{align}
K_{\theta} = K_b \cos^2 \theta_{ba} + K_a \sin^2 \theta_{ba}. 
\end{align}
In the superconducting state, the change in Knight shift includes contributions from the decrease in spin susceptibility $K_{\rm spin}$ as well as the temperature dependence of the Knight shift in the normal state $K_{\rm N}$ and the contribution of the superconducting diamagnetism $K_{\rm dia}$.
Then, all Knight-shift contributions in $H$ along the $i$ direction can be described as
\begin{equation}
K_i(T) = K_{\text{spin},i}(T) + K_{\text{dia},i}(T) + K_{\text{N},i}(T).
\end{equation}
In this study, the temperature variation of $K_{\text{N},i}(T)$ is negligible compared to the decrease in the superconducting state and can thus be ignored.
In addition, for small angles like 10 degrees, the angular variation of $K_{\text{dia},i}$ is also negligible, and thus, we use the value for the $b$ axis $K_{\text{dia},b} = -(H_{c1,b}/H)\times[\ln(\beta \lambda_d/\sqrt{e\xi_a\xi_c})/\ln \sqrt{\kappa_a\kappa_c}]$\cite{deGennes} = $-0.078$\% at the lowest temperature and 0.66 T.
Here, $\xi_i$ is the Ginzburg–Landau (GL) coherence length along $i$ axis, $\beta$ is a factor that depends on the vortex structure and is 0.38 for the triangular vortex lattice, $\lambda_d$ is the distance between the vortices and is calculated using the relation $\phi_0 = \frac{\sqrt{3}}{2}\lambda_d^2(\mu_0H_{\rm ext})$, and $\kappa_i$ is the GL parameter along $i$ axis. 
We used the SC parameters reported by Paulsen $et~al$\cite{C.Paulsen_PRB_2021} for the estimation.
Thus, the decrease in the spin susceptibility along the $a$ axis in the superconducting state, $\Delta K_{\text{spin},a}$, is given by
\begin{align}
\Delta K_{\text{spin},a} = \frac{1}{\sin^2 \theta_{ba}}\{K_\theta(1.6~\text{K}) - K_\theta(0.1~\text{K}) - K_{\text{dia}}) \notag \\
- [K_b(1.6~\text{K}) - K_b(0.1~\text{K}) - K_{\text{dia}}] \cos^2 \theta_{ba}\}.
\label{eq}
\end{align}
Note that $K_{\text{spin},a}$ = 36.1\% just above $T_{\rm SC}$\cite{H.Fujibayashi_JPSJ_2022}.
Using eq.\eqref{eq}, we evaluate that $\Delta K_{\text{spin},a}$ estimated from $\theta_{ba} = 10^\circ$ is 3.0 $\pm$ 0.9\% and estimated from $\theta_{ba} = 12.5^\circ$ is 2.5 $\pm$ 0.8\%.
For $\theta_{ba} = 12.5^\circ$, we adopted the value of the lower $K$ peak in the double-peak structure.
Both values reach almost 3 \%, which is significantly larger than the previously reported value of 0.01\% for the early-stage samples\cite{H.Fujibayashi_JPSJ_2022}.
The estimated reduction for the ultraclean samples at the lowest temperatures is around 5\%\cite{H.Matsumura_JPSJ_2023}.
Considering that the reduction observed for the $H \parallel b$ and $H \parallel c$ axes in the early-stage sample was about half of the reduction observed in the ultraclean samples\cite{H.Fujibayashi_JPSJ_2022,H.Matsumura_JPSJ_2023}, the results in this study are consistent with those in the ultraclean sample.
This indicates that the spin susceptibility along the $a$ axis also decreases in the early-stage samples.
The lack of Knight-shift reduction in the previous studies may be due to the missing signal from the superconducting region.
The signal observed in the previous study may originate from the residual part of the non-superconducting region.
In the previous report\cite{H.Fujibayashi_JPSJ_2022}, the NMR spectrum for $H \parallel a$ did not show broadening below $T_{\rm SC}$, whereas in this study, as shown in Fig.~\ref{fig.2}(h), clear broadening was observed.
This difference also suggests that the signals detected in the previous and current measurements originate from different sample regions.
Although the $a$-axis field component in both measurements is similar, it is considered that the signal from the superconducting region becomes stronger by using the field tilting method.
It is noted that, in the previous measurements\cite{H.Fujibayashi_JPSJ_2022}, we observed a small decrease in the Knight shift just below $T_{\rm SC}$ due to the contribution from bulk superconducting diamagnetism.
This observation indicates that, in the previous experiment, the effects of RF heating were minimal, and the superconducting state was indeed measured.

\begin{figure}[tb]
\centering
\includegraphics[width=0.5\textwidth]{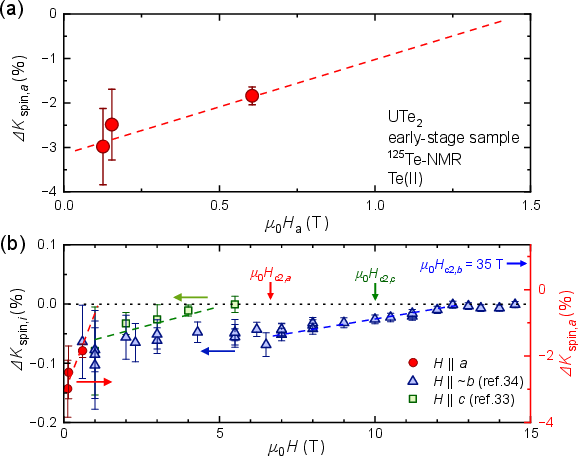}
\caption{
(Color online) (a)Projected magnetic field along the $a$ axis ($\mu_0 H_a$) dependence of the $a$-axis component of decrease in the spin part of Knight shift.
The broken line is a guide for the eye.
(b)Magnetic field dependence of the decrease in the spin part of Knight shift for each axis ($\Delta K_{\textrm{spin},i}$, $i = a,b$, and $c$).
The data for $H \parallel b$\cite{G.Nakamine_JPSJ_2021} and $H \parallel c$\cite{G.Nakamine_PRB_2021} are taken from the reference.
The broken lines are guides for the eye.
The arrows indicate $H_{\rm c2}$ for each axis.
}
\label{fig.3}
\end{figure}

Furthermore, the estimation of the $a$-axis component from the results at 3~T indicates a reduction of 1.8 $\pm$ 0.2\% in $\mu_0 H_a = 3\times \sin10.5^\circ = 0.55$~T.
The $a$-axis projected magnetic field dependence of the decrease in the $a$-axis
spin susceptibility is shown in Fig.~\ref{fig.3}(a).
The reduction in the $a$-axis spin susceptibility is suppressed by increasing the magnetic field.
The extrapolation suggests that the spin susceptibility would be unchanged in the superconducting state at $\mu_0 H_a \sim$ 1.5 T.
Considering that the upper critical field $H_{\rm c2}$ in the $a$-axis direction is approximately 7 T for the early-stage sample\cite{D.Aoki_JPSJ_2019_2}, the suppression of the spin susceptibility reduction can be attributed to the rotation of superconducting spins.

Near zero magnetic field, it is considered that the most promising superconducting symmetry of UTe$_2$ is $A_u$ with an irreducible representation of $D_{2h}$ point group that has $d$-vector components, which is perpendicular to superconducting spin, in all directions, similar to the B phase of superfluid $^3$He\cite{A.J.Leggett_RevModPhys_1975}.
When a magnetic field is applied, the superconducting spins polarize and align along the field direction.
Such behavior is theoretically predicted for spin-triplet superconductors\cite{K.Machida_PRB_2023} and has been experimentally observed in UTe$_2$ for $H \parallel b$ and $H \parallel c$\cite{G.Nakamine_PRB_2021,G.Nakamine_JPSJ_2021,K.Kinjo_PRB_2023}, as well as in UPt$_3$\cite{H.Tou_PRL_1998}.
Our study is the first example to observe the rotation of superconducting spins along all crystal axes in a spin-triplet superconductor.
The complete rotation fields in the early-stage sample are 12 T for $H \parallel b$, 5 T for $H \parallel c$\cite{G.Nakamine_PRB_2021}, and about 1.5 T for $H \parallel a$, as shown in Fig.~\ref{fig.3}(b).
These results demonstrate the anisotropy of the spin rotation fields among the crystal axes, and the $d$-vector pinning interaction is not so large.
The experimental results that the magnetic field of superconducting spin rotation is lowest in the magnetic easy $a$ axis and that superconducting spin rotation does not occur until a large magnetic field in the magnetic hard $b$ axis suggests a close relationship between the magnetic properties of the normal state and superconducting spin rotation in spin-triplet superconductors.

In conclusion, we have re-measured the $a$-axis component of the spin susceptibility in the early-stage sample of the spin-triplet superconductor UTe$_2$.
Instead of directly measuring the $a$ axis spectrum with a broad linewidth and weak signal intensity, we measured the spin susceptibility for the $b$ axis and the field tilted by 10 degrees from the $b$ axis towards the $a$ axis.
This approach allowed us to accurately determine the spin susceptibility component along the $a$ axis.
Our results indicate that the spin susceptibility along the $a$ axis decreases by approximately 3\% in a low magnetic field of $\mu_0 H_a \sim 0.1$ T.
This suggests that even in the early-stage sample, the $a$-axis spin susceptibility decreases similarly to that in the ultraclean samples, indicating the existence of $d$-vector component along the $a$-axis.
The absence of Knight shift reduction in previous measurements is likely due to the missing signal from the superconducting region and to detecting the residual signal from the non-superconducting region instead.
Moreover, we have shown that the decrease in the spin susceptibility along the $a$ axis is immediately suppressed with increasing the $a$-axis magnetic field, and it is anticipated to be unchanged in the superconducting state at around 1.5 T.
Our findings reveal that the spin-rotation field in the superconducting state has anisotropy among the crystal axes, which is linked with the normal-state magnetic properties, and that the extremely high $H_{\rm c2}$ observed in all axes is the consequence of the field-polarized triplet pair.
Our results advance the understanding of the unique superconducting properties and impurity effects in spin-triplet superconductors, contributing significantly to the field.

\section*{acknowledgments}
We acknowledge H. Tou, Y. Yanase, and J.-P. Brison for fruitful discussion. 
This work was supported by Grants-in-Aid for Scientific Research (KAKENHI Grant No. JP20KK0061, No. JP20H00130, No. JP21K18600, No. JP22H04933, No. JP22H01168, No. JP23H01124, No. JP23K22439 and No. JP23K25821) from the Japan Society for the Promotion of Science, by JST SPRING(Grant No. JPMJSP2110) from Japan Science and Technology Agency, by research support funding from The Kyoto University Foundation, by ISHIZUE 2024 of Kyoto University Research Development Program, and by Murata Science and Education Foundation.
In addition, liquid helium is supplied by the Low Temperature and Materials Sciences Division, Agency for Health, Safety and Environment, Kyoto University.

\end{document}